\documentclass[preprint,prd,nofootinbib]{revtex4-1}
\usepackage{color}
\usepackage{graphicx}
\usepackage{epsfig}
\usepackage{subfig}
\usepackage{ctable}
\usepackage{hyperref}
\usepackage{graphicx}
\usepackage{dcolumn}
\usepackage{bm}

\begin{document}

 \begin{center}
     \Large{\textbf{  Anisotropic Stars in the Non-minimal $Y(R)F^2$ Gravity  }} \\[0.3cm]

     \large{\"{O}zcan SERT$^{\rm a}$,  Fatma \c{C}EL\.{I}KTA\c{S}$^{\rm a}$},
        Muzaffer ADAK$^{\rm b}$
        \\[0.3cm]

        \small{
            \textit{$^{\rm a}$ Department of Mathematics, Pamukkale University, 20017 Denizli, Turkey}}

        \small{
            \textit{$^{\rm b}$ Department of Physics, Pamukkale University, 20017 Denizli, Turkey}}
    \end{center}


\hrule \vspace{0.2cm}
\noindent \small{\textbf{Abstract}

We investigate anisotropic compact stars in the non-minimal $Y(R)F^2$ model of  gravity which couples an arbitrary function of curvature scalar $Y(R)$ to the electromagnetic field invariant $F^2$. After we obtain exact anisotropic solutions to the field equations of the model,  we apply the continuity conditions  to the solutions at   the
boundary of the star. Then we find the mass, electric charge, and  surface gravitational redshift by the   parameters of the model and radius of the star.

\section{Introduction}

Compact  stars are the best  sources to test a theory of gravity under the extreme cases with strong fields. Although they are generally considered as isotropic, there are important reasons to take into account anisotropic compact stars which have different radial and tangential pressures. First of all, the anisotropic spherically symmetric compact stars can be more stable than the isotropic ones \cite{Dev}. The core region of the compact stars with very high nuclear matter density becomes more realistic in the presence of anisotropic pressures \cite{Ruderman,Canuto}. Moreover the phase transitions \cite{Sokolov}, pion condensations \cite{Sawyer} and the type 3A superfluids \cite{Kippenhahm} in the cooling  neutron matter core can lead to anisotropic pressure distribution. Furthermore, the mixture of two perfect fluid can generate anisotropic fluid \cite{Letelier}. Anisotropy can be also  sourced by the rotation of the star \cite{Herrera,Bayin,Silva}. Additionally,  strong magnetic fields may lead to anisotropic pressure components in the compact stars\cite{Mak2002}. Some analytic solutions of anisotropic matter distribution were studied in Einsteinian Gravity \cite{Herrera,Bayin,Silva,Mak2002,Delgaty,Harko2002,Mak2003,Bowers,Mak20022,Durgapal1985}. Recently it was shown that the "scalarization" can not arise without anisotropy and the anisotropy range can be determined by observations on binary pulsar in the Scalar-Tensor Gravity and General Relativity \cite{Silva}. 
The anisotropic star  solutions  
in $R^2$ gravity can shift the mass-radius curves to the region given by observations 
\cite{Folomeev}. It is interesting to note that anisotropic compact stars were  investigated in Rastall theory and found exact solutions   which  permit the formation of super-massive star \cite{Salako2018}.

Additionally,  the presence of a constant  electric charge on the surface of   compact stars may increase the stability \cite{Stettner} and protect them from collapsing  \cite{Krasinski,Sharma}.
The charged fluids can be  described  by the minimally coupled Einstein-Maxwell field equations.  
An exact isotropic  solution of the Einstein-Maxwell theory were found by Mak and Harko describing  physical parameters of a   quark star with the  MIT bag equation of state under the existence of conformal motions \cite{mak-harko-2004}. Also,    the  upper and lower limits for the basic physical quantities  such as mass-radius ratio, 
redshift were derived for charged compact stars \cite{Mak2001}
and 
for anisotropic stars \cite{Bohmer2006}.
A regular charged solution of the field equations which satisfy physical conditions was found in \cite{Krori} and the constants of the solution were fixed in terms of mass, charge and radius \cite{Junevicus}. Later the solutions  were extended to the charged anisotropic fluids \cite{Varela,Rahaman}.
Also,  anisotropic charged fluid spheres were studied in $D$-dimensions \cite{Harko2000}.

In the investigation of compact stars,  one of the most important  problem is the mass discrepancy between the predictions of nuclear theories \cite{Xu2010,Glendenning1991,Panda2004,Page2006} and  neutron star observations. 
The resent observations such as $1.97M_\odot $ of the neutron star PSR J1614-2230 \cite{Demorest2010},  $2.4M_\odot $  of  the neutron stars B1957+20 \cite{Kerkwijk2011} and 4U 1700-377 \cite{Clark2002} or $2.7M_\odot $ of  J1748-2021B \cite{Freire2008} can not be explained  by using any soft  equation of state in Einstein's gravity \cite{Xu2010,Demorest2010,Wen2012}. 
  Since each different approach which solves this problem leads to different maximal mass, we need to more reliable  models which satisfy observations and give  the correct  maximum mass limit of compact stars.

On the other hand, the observational problems such as dark energy and dark matter  \cite{Overduin,Baer,Riess,Perlmutter,Knop,Amanullah,Weinberg,Schwarz}  at astrophysical scales have caused to search new modified  theories of gravitation such as $f(R)$ gravity \cite{Capozz2011,Nojiri2011,Nojiri2007,Capozz2010,Capozz2003,Joyce,Capozz2012}.
As an alternative approach, the $f(R)$ theories of gravitation can explain the inflation and cosmic acceleration  without exotic fields and satisfy the cosmological observations \cite{Nojiri2003,Nojiri20032}.  
Therefore, in the strong gravity regimes such as inside the compact stars,  $f(R) $ gravity models       can be considered to describe   the more massive stars \cite{Capozziello2016,Astashenok2014} .
Furthermore, the  strong magnetic fields \cite{Astashenok2015} and electric fields \cite{Jing2015}  can   increase the mass of  neutron stars  in the framework of  $f(R)$ gravity.

 On the other hand, in the presence of the strong  electromagnetic fields, the Einstein-Maxwell theory can also be modified. The first modification is the minimal coupling between $f(R)$ gravity and Maxwell field as the $f(R)-$Maxwell gravity
which has only the spherically symmetric static solution
\cite{Dombriz2009}. Therefore we consider the more general modifications  such as $Y(R)F_{ab}F^{ab}$ which allow a wide range  of solutions \cite{AADS,dereli3,Sert12Plus,Sert13MPLA,bamba1,bamba2,Sert-Adak12,dereli4,Turner,Mazzi,Campa}.

 
 A similar kind of such a modification which is $ R_{abcd}F^{ab} F^{cd} $   firstly 
 was defined   by Prasanna \cite{Prasanna} and found   a criterion for null  electromagnetic fields in conformally flat space-times.
Then the most general possible  invariants which involves electromagnetic and gravitational fields (vector-tensor fields)   including $RF_{ab}F^{ab}$ term    were studied in \cite{Horndeski} and  the spherically symmetric static solutions  were obtained  \cite{Muller-Hoissen1988}  for  the unique composition of such couplings.
Also  such non-minimal modifications   were  derived by 
   dimensional  reduction of the five dimensional  Gauss-Bonnet  action \cite{Muller-Hoissen1988-2,Muller-Hoissen1988-3} and  $R^2$  gravity action \cite{Buchdahl1979,Dereli1990}. 
   These invariant terms were also obtained by Feynman diagram method from vacuum polarization of the photon  in the
   weak gravitational field limit \cite{Drummond1980}. The general form of the couplings 
    such as $R^nF^2$ 
 were used to explain    the  seed magnetic fields in  the inflation and 
 the production of  the primeval magnetic flux in the universe  \cite{Turner,Campa,Kunze,Mazzi,dereli4,bamba1}.

 Thus it is natural to consider  the more general modifications  which couple  a function of the Ricci scalar with Maxwell invariant as $Y(R)F^2$ form  inside the  strong  electromagnetic and gravitational fields such as  compact astrophysical objects.
 The more general modifications have  static spherically symmetric solutions  \cite{Sert12Plus,dereli3,dereli4,Sert13MPLA} 
to describe the flatness of velocity curves of galaxies, cosmological solutions to describe accelerating expansion of the universe \cite{Sert-Adak12,AADS,bamba1,bamba2} and regular black hole solutions \cite{Sert2016regular}.
Furthermore, the charged, isotropic stars and radiation fluid  stars can be described by the non-minimal couplings \cite{Sert2017,Sert2018}.
Therefore in this study, we investigate the anisotropic compact stars in the non-minimal $Y(R)F^2$ model  and find a family of exact analytical solutions. Then we obtain the total mass, total charge and gravitational surface redshift by the parameters of the model and the boundary radius of the star.

\section{The Model for Anisotropic Stars} \label{model}

We will obtain the field equations of our model for the anisotropic stars by
varying the action integral with respect to independent variables;
the orthonormal co-frame 1-form $e^a$, the Levi-Civita connection
1-form $\omega_{ab}$, and the electromagnetic potential 1- form
$A$,
\begin{equation}\label{action}
I  = \int_M{\left\{   \frac{1}{2\kappa} R*1 - \epsilon_0 Y(R) F\wedge *F + \frac{2}{c}A\wedge J + L_{mat} + \lambda_a\wedge T^a  \right\} }
\end{equation}
where $\kappa$ is the coupling constant, $\epsilon_0$ is the permittivity of free space, $\wedge$ and $*$ denote the exterior product and the Hodge map of the exterior algebra, respectively, $R$ is the curvature scalar, $Y(R)$ is a function of $R$ representing the non-minimal coupling between gravity and electromagnetism, $F$ is the electromagnetic 2-form, $F=dA$, $J$ is the electromagnetic current 3-form,
$L_{mat}$ is the matter Lagrangian 4-form, $T^a := de^a + \omega^a{}_b \wedge e^b$ is the torsion
2-form, $\lambda_a$ Lagrange multiplier 2-form constraining
torsion to zero and $M$ is the differentiable four-dimensional manifold whose orientation is set by the choice $*1=e^0 \wedge e^1 \wedge e^2 \wedge e^3$.

In this study, we  use  the SI units differently from the previous papers \cite{Sert2017,Sert2018}. We write $F = -E_i e^{0i} + \frac{c}{2}\epsilon_{0ijk}B^k e^{ij}$, where $c$ is the light velocity, $E_i$ is electric field and $B_i$ is magnetic field under the assumption of the Levi-Civita symbol $\epsilon_{0123}=+1$. We adhere the following convention about the indices; $a,b,c = 0,1,2,3$ and $i,j, k = 1,2,3$.  We also define $J=-c \rho_e *e^0 + J_i *e^i$ where $\rho_e$ is the  charge density and $J_i$ is the  current density. We assume that the co-frame variation of the matter sector of the Lagrangian produces the following energy momentum 3-form
 \begin{eqnarray}
   \tau_a^{mat} := \frac{\partial L_{mat}}{\partial e^a}  
   = (\rho c^2 + p_t )u_a *u + p_t *e_a + (p_r - p_t ) v_a *v
 \end{eqnarray}
where $u = \delta_a^0 e^a $ is a time-like 1-form, $v= \delta_a^1 e^a$ is a space-like 1-form, $p_r$ is the radial component of pressure orthogonal to the transversal pressure $p_t$, and $\rho$ is the  mass density for the anisotropic matter in the star. In this case it must be $\kappa = 8\pi G / c^4$ for the correct Newton limit. The non-minimal coupling function $Y(R)$ is dimensionless. Consequently every term in our Lagrangian has the dimension of $(energy)(length)$. Finally we notice that the dimension of $\lambda_a$ must be $energy$ because torsion has the dimension of $length$. 

After substituting the connection varied equation into the co-frame varied one, we obtain the modified Einstein equation for our model
 \begin{eqnarray} \label{gfe}
   - \frac{1}{2 \kappa}  R^{bc} \wedge *e_{abc}  &=& \epsilon_0 Y(\iota_a F \wedge *F
   - F \wedge \iota_a *F) + \epsilon_0 Y_R F_{bc} F^{bc}*R_a \nonumber \\
   & & + \epsilon_0 D[\iota^b \; d(Y_R F_{bc} F^{bc})] \wedge *e_{ab} + \tau_a^{mat}    \  , \hskip 1 cm
 \end{eqnarray}
where $\iota$ denotes the interior product of the exterior algebra satisfying the duality relation, $\iota_b e^a = \delta^a_b$, through the Kronecker delta, $R^a{}_b := d \omega^a{}_b + \omega^a{}_c \wedge \omega^c{}_b $ is the Riemann curvature 2-form, $ R^a:= \iota_b R^{ba} $ is the Ricci curvature 1-form, $R := \iota_a R^a $ is the curvature scalar, $Y_R :=dY/dR$, $e^{ab \cdots} := e^a \wedge e^b \wedge \cdots $ and $F_{bc}:=\iota_c \iota_b F$. As the left hand side of the equation (\ref{gfe}) constitutes the Einstein tensor, the right one is called the total energy momentum tensor for which the law of energy-momentum conservation is valid. For details one may consult Ref. \cite{Sert2017}.

The variation  of the action according to the electromagnetic potential $A$  yields the modified Maxwell equation
 \begin{eqnarray}
     \epsilon_0 d(*Y F) &=& \frac{J}{c}  \; .   \label{maxwell1}
 \end{eqnarray}
We have also noticed that  the Maxwell 2-form   is closed $dF=0$  from the Poincar\'e lemma, since it  is exact form $F=dA$. Thus,  the two field equations (\ref{gfe}) and (\ref{maxwell1}) define our model and we will look for solutions to them under the condition
 \begin{eqnarray}
  \epsilon_0 Y_R F_{bc} F^{bc} = - \frac{k}{ \kappa}  \label{YRFk}
 \end{eqnarray}
which removes  the potential  instabilities from the higher order derivatives for  the non-minimal  theory. Here the non-zero  $k$  is a dimensionless constant. The case of $k=0$ leads to $Y(R)= constant$ corresponding to the minimal Einstein-Maxwell theory which will be considered as the exterior vacuum solution  with $R=0$.  We will see from solutions of the model that the total mass and total charge of the anisotropic star are critically  dependent on  the parameter $k$. The additional features of the constraint (\ref{YRFk}) may be found in \cite{Sert2017}. We also notice that the trace of the modified Einstein equation (\ref{gfe}), obtained by multiplying with $e^a \wedge$, produces an explicit relation between the Ricci curvature scalar, energy density  and pressures  
 \begin{eqnarray}
   (1- k) R = \kappa (\rho c^2 - p_r-2p_t) \;. \label{trace}
 \end{eqnarray}

\section{ Static spherically symmetric Anisotropic solutions } \label{model}

We propose the following metric for the static spherically symmetric spacetime and the Maxwell 2-form for the static electric field parallel to radial coordinate in (1+3) dimensions
 \begin{eqnarray}
  ds^2 &=& -f^2(r) c^2 dt^2+g^2(r)dr^2+r^2d\theta^2 \label{metric1} +r^2\sin^2\theta d\phi^2 \; ,\\
     F &=& E(r)e^1\wedge e^0 \; ,
 \end{eqnarray}
where $f$ and $g$ are the metric functions and $E$ is the electric field in the radial direction which all three functions  depend  only on  the radial coordinate $r$.

The integral of the electric current density 3-form $J$ sourcing the electric field gives rise to the electric charge inside the volume $V$ surrounded by the closed spherical surface $\partial V$ with radius $r$
 \begin{eqnarray}
 q(r):=\frac{1}{c}\int_V{J} = \epsilon_0 \int_V{d(*YF)} = \epsilon_0 \int_{\partial V} {*YF} = 4\pi \epsilon_0 r^2 Y E \, . \label{charger}
 \end{eqnarray}
Here we used the Stoke's theorem. The components of the modified Einstein equation (\ref{gfe}) reads three coupled nonlinear differential equations
  \begin{eqnarray}
  \frac{1}{\kappa} \left( \frac{2g'}{g^3 r} + \frac{g^2 - 1}{g^2 r^2} \right) = \epsilon_0 YE^2 - \frac{k }{\kappa}\left( \frac{f''}{fg^2} - \frac{f'g'}{fg^3} + \frac{2f'}{f g^2 r}\right) + \rho c^2 \; , \label{eqns1}\\
  \frac{1}{\kappa } \left(-\frac{2f'}{f g^2 r} + \frac{g^2 - 1}{g^2 r^2} \right) = \epsilon_0 YE^2 + \frac{k }{\kappa}\left( -\frac{f''}{fg^2} + \frac{f'g'}{fg^3} + \frac{2g'}{g^3 r}\right)- p_r \; , \\
 \frac{1}{\kappa} \left(\frac{f''}{f g^2} - \frac{f'g'}{f g^3} + \frac{f'}{fg^2 r} - \frac{g'}{g^3 r} \right) = \epsilon_0 YE^2 - \frac{k }{\kappa}\left( -\frac{f'}{fg^2 r} +\frac{g'}{g^3 r} + \frac{g^2 - 1}{g^2 r^2} \right) + p_t \; ,
\end{eqnarray}
where prime stands for derivative with respect to $r$. We obtain one more useful equation by taking the covariant exterior derivative of (\ref{gfe})
 \begin{eqnarray}
p_r' +  \frac{f'(\rho c^2 + p_r)}{f} +  \frac{2(p_r - p_t)}{r} = 2\epsilon_0 (YE)' E + \frac{4 \epsilon_0 YE^2}{r} \, .
 \end{eqnarray}
In what follows we assume  the linear equation of state in the star 
 \begin{eqnarray}
   p_r = \omega \rho c^2 \label{eqnofstate}
 \end{eqnarray}
where $\omega$ is a constant in the interval $0<\omega \leq 1 $. Finally we calculate the crucial constraint (\ref{YRFk}) 
 \begin{eqnarray}
   \frac{dY}{dR}=\frac{k}{2\epsilon_0\kappa E^2 } \; , \label{cond2}
 \end{eqnarray}
where the Ricci curvature scalar is
\begin{eqnarray}\label{R}
 R = \frac{2}{g^2 }\left( -\frac{f''}{f} + \frac{f'g'}{fg} -\frac{2f'}{fr} + \frac{2g'}{gr} + \frac{g^2-1}{ r^2} \right) \; .
 \end{eqnarray}

 \subsection{Exact solutions with conformal symmetry} \label{model}
We will look for solutions to these differential equations  (\ref{eqns1})-(\ref{cond2}) assuming that the metric (\ref{metric1})  admits   a one-parameter group of
conformal motions, since inside of stars can be described
 by using this symmetry \cite{mak-harko-2004,Herrera1,Herrera2,Herrera3}. The symmetry  is  obtained by taking  Lie derivative
 of the metric tensor $\mathbf{g}_{ab}$  with respect to the vector field $\xi$, $L_\xi \mathbf{g}_{ab}= \frac{\phi_0}{g(r)} \mathbf{g}_{ab}$,   for  the arbitrary metric function $g(r)$ and the following metric function $f(r)$ 
 \begin{eqnarray}\label{f}
 f^2(r) = a^2 r^2
 \end{eqnarray} 
with arbitrary constants  $a$ and $\phi_0$. Here we consider  the  metric function $g(r) $ 
 \begin{eqnarray}\label{g}
  g^2(r) = \frac{3}{1+br^\alpha} \; ,
 \end{eqnarray}
 inspired by \cite{mak-harko-2004},
 where $b$ and $\alpha$ are arbitrary parameters.
 With these choices, the curvature scalar (\ref{R}) is calculated as
 \begin{eqnarray}
 R = -b(\alpha + 2) r^{\alpha-2} \, . \label{cursca}
 \end{eqnarray}
 We notice that it must be  $\alpha > 2$ and $b \neq0 $ in order for that the curvature scalar is nonzero and regular at the origin. If $b = 0$, then the curvature scalar $R$ becomes zero and this leads to constant $Y(R)$ in which case the model reduces to the minimal Einstein-Maxwell theory. 
 
Then the system of equations   (\ref{eqns1})-(\ref{cond2})  has 
 the following solutions for the metric functions (\ref{f}), (\ref{g}) and the anisotropic pressure $p_r = \omega \rho  c^2$
 \begin{eqnarray}
 & &   \rho(r)=\frac{(k  + 1)[2- br^\alpha (\alpha-2)  ]}{3\kappa c^2 r^2 (\omega +1)} \, , \label{rho1} \\
\\
 & &  p_t(r)=\frac{br^\alpha [(k  +1)(\alpha -2 ) -X ]}{3\kappa r^2 (\omega +1)}   + \frac{(k+1)(1- \omega )}{3\kappa r^2(\omega +1)}\, , \label{pt1} \\
 & &   E^2(r)= \frac{2\omega(k+1)}{3\kappa \epsilon_0 c_0 r^2 (\omega+1)} \left[1+\frac{X br^\alpha}{4\omega(k  + 1)}\right]^{1+ \frac{3k (\omega+1)(\alpha^2-4)}{\alpha X}} \, , \label{E1}\\
  & & Y(r)=c_0\left[1+\frac{X br^\alpha}{4\omega(k  + 1)}\right]^{-\frac{3k (\omega+1)(\alpha^2-4)}{\alpha X}} \, , 
 \end{eqnarray}
 where  the composite function $ Y(R(r))$  have obtained  in terms of $r$ as  $Y(r)$ and we have defined 
 \begin{eqnarray}
 X = k  \omega(\alpha + 4) + \alpha(3k - 2\omega)-2(\omega+3) \, .
 \end{eqnarray}

 After obtaining $r(R)$ from (\ref{cursca})
\begin{eqnarray}
r = \left[\frac{-R}{b(\alpha  + 2)}\right]^{1/(\alpha - 2)} \label{rofR}
\end{eqnarray}  we rewrite explicitly the non-minimal coupling function in terms of $R$
 \begin{eqnarray}\label{Y1}
 Y(R) = c_0\left[1+ \frac{b X}{4\omega(k+1)}\left(\frac{-R}{\alpha b +2b}\right)^{\alpha/(\alpha-2)} \right]^{-\frac{3k(\omega+1)(\alpha^2-4)}{\alpha X}} .
 \end{eqnarray}
   Since the exterior vacuum region is described by Reissner-Nordstr\"om metric satisfying $R=0$, the non-minimal coupling function becomes $Y(R) = c_0 $. Therefore we can fix $c_0=1$ without loss of generality.
   Then the corresponding model becomes
  
  \begin{eqnarray}\label{theory}
  L =    
  \frac{1}{2\kappa^2} R*1
  -       \epsilon_0 \left[1+ \frac{b X}{4\omega(k+1)}\left(\frac{-R}{(\alpha+2) b }\right)^{\frac{\alpha}{\alpha-2}} \right]^{-\frac{3k(\omega+1)(\alpha^2-4)}{\alpha X}} \hskip-0.7 cm F\wedge *F  + 2A\wedge J + L_{mat}  + \lambda_a\wedge T^a . \ \ \ \ \ \ 
  \end{eqnarray} 
admitting    the interior  metric
 \begin{eqnarray}\label{metricin}
 ds^2 _{in}= - a^2r^2 c^2 dt^2 + \frac{3}{1+br^\alpha} dr^2 + r^2 d\Omega^2 \;
 \end{eqnarray}
and the the Maxwell 2-form 
 \begin{eqnarray}
 F = \frac{q(r)}{4 \pi \epsilon_0 Y r^2} e^1 \wedge e^0
 \end{eqnarray}
 in the interior of star, where $q(r)$ is obtained from  (\ref{charger})  as
 \begin{eqnarray}\label{q}
 q^2(r)=\frac{32\pi^2\epsilon_0\omega  (k + 1)r^2}{3\kappa  (\omega +1)}\left[1+\frac{X br^\alpha}{4\omega(k + 1)}\right]^{1-\frac{3k(\omega+1)(\alpha^2-4)}{\alpha X}} \,  .
 \end{eqnarray}
 
On the other hand  at the exterior, the model admit the Reissner-Nordstr\"om metric
 \begin{eqnarray}\label{metricext}
 ds^2_{out} = -\left( 1- \frac{2GM}{c^2r} + \frac{\kappa Q^2}{(4 \pi)^2 \epsilon_0 r^2} \right) c^2 dt^2 + \left( 1-\frac{2GM}{c^2r} + \frac{\kappa Q^2}{(4 \pi)^2 \epsilon_0 r^2} \right)^{-1} dr^2 + r^2d\Omega^2 \;
 \end{eqnarray}
and the Maxwell 2-form\footnote{Here we  use  the SI units differently from  \cite{Sert2017,Sert2018} in  which leads to $F= \frac{Q}{r^2} e^1\wedge e^0$ and $\frac{\kappa Q^2}{r^2} $ term in the metric.}
 \begin{eqnarray}
 F = \frac{Q}{4 \pi \epsilon_0 r^2} e^1 \wedge e^0
 \end{eqnarray}
  which represents the electric field in the radial direction, $ \vec{E} = {\hat{r} Q}/{4 \pi \epsilon_0 r^2} $,
  where $Q$ is the total electric charge of the star which  is obtained by writing $r=r_b$ in (\ref{q}).
   We will be able to determine some of the parameters from the matching and the continuity conditions, and the others from the observational data.

\subsection{Matching conditions} \label{model}

We will match the interior metric (\ref{metricin}) and the exterior metric (\ref{metricext}) at the boundary of the star $r=r_b$ for continuity of the gravitational potential,
 \begin{eqnarray}
  a^2 r_b ^2 &=& 1 - \frac{2GM}{c^2 r_b} + \frac{\kappa Q^2}{(4\pi)^2 \epsilon_0 r_b^2} \, , \\
  \frac{3}{1 + br^\alpha_b} &=& \left(1 - \frac{2GM}{c^2 r_b} + \frac{\kappa Q^2}{(4\pi)^2 \epsilon_0 r_b^2}\right)^{-1} \, .
 \end{eqnarray}
These equations are solved for $a$ and $b$ appeared in the interior metric functions
 \begin{eqnarray}
    & & b = \frac{2r_b^2 - 6GMr_b/c^2 + 3 \kappa Q^2/(4\pi)^2\epsilon_0}{r_b^{2+\alpha}} \, ,  \label{bMQ} \\
    & & a^2 =  \frac{1}{r_b^2} - \frac{2GM}{c^2 r_b^3} + \frac{\kappa Q^2}{(4\pi)^2 \epsilon_0 r_b^4} \, .
\end{eqnarray}
Vanishing of the radial pressure at the boundary, $p_r(r_b)= \omega  c^2 \rho(r_b)  = 0$ in (\ref{rho1}), determines the parameter $b$
 \begin{eqnarray} \label{bpara}
   b=\frac{2}{(\alpha-2)r_b^\alpha} \; .
 \end{eqnarray}
 
 \begin{figure}[t]{}
	\centering
	\subfloat[ $ p_r(r) $  ]{ \includegraphics[width=0.5\textwidth]{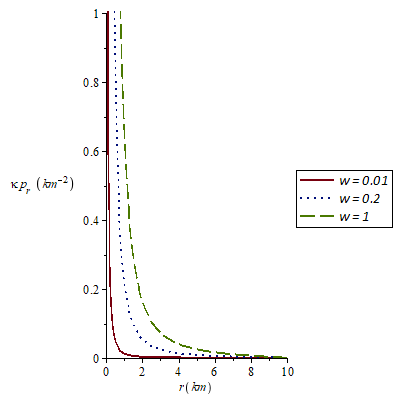}}
	\subfloat[ $p_t(r) $ ]{ \includegraphics[width=0.5\textwidth]{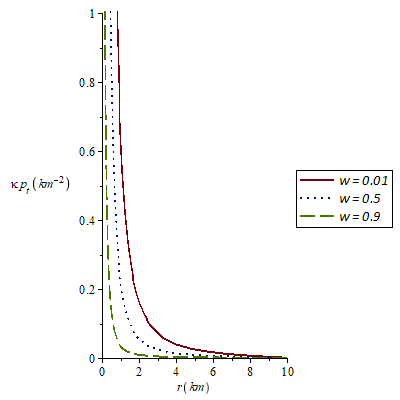} }\\
	\parbox{6in}{\caption{{{\small{   The radial (a)  and tangential (b) pressures as a function of the radial distance $r$  for $k=1$, $\alpha=4$, $r_b=10 km$ and some different $\omega $ values. (The energy density $ c^2\rho $   is proportional to the radial pressure  as $p_r =\omega c^2\rho$ )
	}}}}}
\end{figure} 
 
 \begin{figure}[h]{}
 	\centering
 	\subfloat[  $\omega=0.1$ ]{ \includegraphics[width=0.5\textwidth]{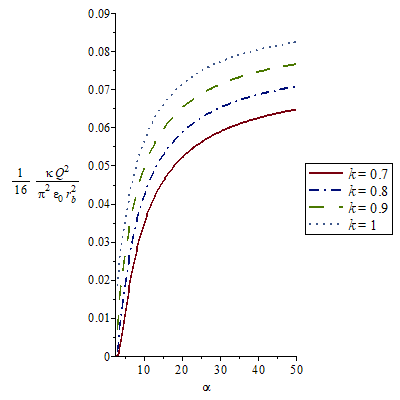} }
 	\subfloat[  $\omega=0.3$]{ \includegraphics[width=0.5\textwidth]{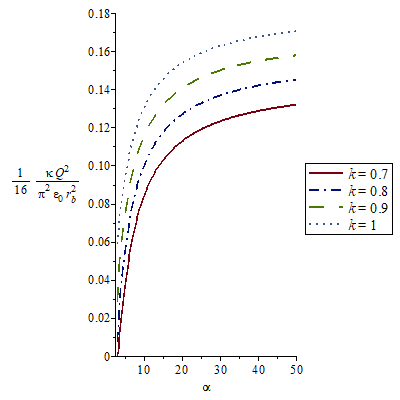}} 
 	\\
 	\parbox{6in}{\caption{{{\small{  The dimensionless quantity which is related with the total charge  $Q$ as a function of the parameter  $\alpha$ for $\omega=0.1$(a) and $\omega=0.3$ (b).  
 	}}}}}
 \end{figure}

\begin{figure}[h]{}
	\centering
	\subfloat[ $\omega=0.15$ ]{ \includegraphics[width=0.5\textwidth]{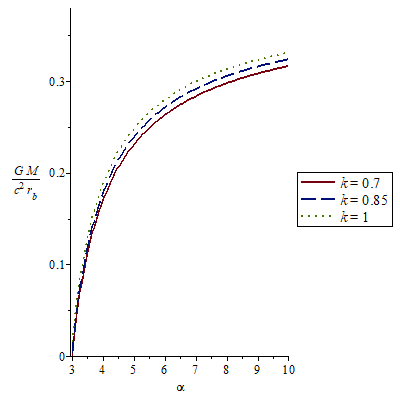} }
	\subfloat[   $\omega=0.5$]{ \includegraphics[width=0.5\textwidth]{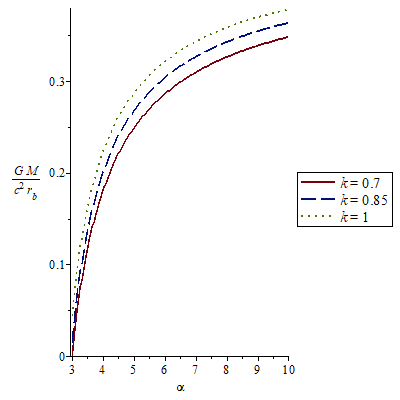}} 
	\\
	\parbox{6in}{\caption{{{\small{  The dimensionless quantity  related with the total  mass as  a function of  $\alpha$ for $\omega= 0.15$ (a) and $\omega =0.5$ (b). 
	}}}}}
\end{figure}

 The behavior of the  pressures and energy density    is given
 in Fig. 1  in terms of the radial distance $r$ for $k=1$. Moreover the decreasing behavior of the quantities does not change for $k\neq 1$.
  We can obtain  an upper bound of the parameter $k$ using  the non-negative tangential pressure $p_t(r)$ in (\ref{pt1}). 
In order to obtain  the bound we need to determine the interval of $\omega$. Since the radial  component of the sound velocity  $ \frac{d p_r}{d\rho} $ is non-negative and should not be bigger than   the square of  light velocity $c^2$ for the normal matter,  the parameter $\omega $ takes values in the range $0\leq \omega \leq 1$. We see from the Fig. 1b that the tangential pressure curves  decrease for  the increasing  $\omega$ values and the minimum curve can be obtained from the case with $\omega = 1$.  Then   we obtain  the  following inequality  from the first part of $p_t(r)$
 \begin{eqnarray}\label{ineq1}
 (k+1)(\alpha-2) -X \geq 0
 \end{eqnarray} 
which turns out to be
   \begin{eqnarray}
   3(1-k)(\alpha+2) \geq 0
   \end{eqnarray} 
     for the case  $\omega=1$, that leads to the minimum $p_t(r) $ function.
   Thus the parameter $k$ must be $k \leq 1$ for the non-negative tangential pressure. On the other hand, the total electric charge  (\ref{Q2}) must be a real valued exponential function,  then  we obtain the inequality 
   \begin{eqnarray}
   1+\frac{2X}{4\omega(k+1)(\alpha-2)} \geq 0
   \end{eqnarray}
   which leads to 
   $k\geq \frac{2}{\alpha}$.  Thus $k$ must take values in this range 
   \begin{eqnarray}
   \frac{2}{\alpha}<k \leq 1.
   \end{eqnarray}
  and  we find the maximum lower bound for the parameter $k$  as $k>\frac{2}{3}$   for $\alpha>3 $ (which leads to positive gravitational redshift).

While the interior side of the star has   the electrically charged  matter distribution,  the outer side  is vacuity. Then the excitation 2-form $\mathcal{G} = YF$  in the interior  becomes the Maxwell 2-form $\mathcal{G} = F$ at the exterior. Therefore the interior electric charge $q(r)$ which  found from $\epsilon_0 c d*YF = J $   must be equal to the total electric charge $Q$  obtained from  $ d* F = 0 $ at the exterior.     Thus the continuity of the excitation 2-form at the boundary leads to $q(r_b)= Q$ as the last matching condition 
 \begin{eqnarray}\label{Q2}
Q^2=\frac{32\pi^2\epsilon_0 \omega  (k +1)r_b^2 }{3\kappa  (\omega +1)}\left[1+\frac{2X}{4\omega(k+1)(\alpha-2)}\right]^{1 -\frac{3k(\omega+1)(\alpha^2-4)}{\alpha X}}
\end{eqnarray}
where we eliminated the parameter $b$ via (\ref{bpara}). The total electric charge  is shown     as a function of $\alpha$ in Fig. 2 for
$\omega =0.1 $ (a) and  $\omega=0.3 $  (b) taking some different $k$ values. We see that the increasing $k$ values increase the total charge values. The substitution of (\ref{bpara}) to (\ref{bMQ}) allows us writing the total mass of the star in terms of its total charge 
\begin{eqnarray}
M = \left(\frac{\alpha -3}{\alpha -2}\right)\frac{c^2 r_b}{3G} + \frac{\kappa c^2 Q^2 }{2(4\pi)^2 \epsilon_0 G r_b} \, . \label{totmass}
\end{eqnarray}
Then from (\ref{Q2}) the total mass becomes
\begin{eqnarray}\label{totmass1}
M=\left(\frac{\alpha -3}{\alpha -2}\right)\frac{c^2 r_b}{3G} +\frac{c^2r_b\omega  (k+1)}{3G (\omega +1)}\left[1+\frac{2X}{4\omega(k+1)(\alpha-2)} \right]^{1-\frac{3k(\omega+1)(\alpha^2-4)}{\alpha X}}.
\end{eqnarray}
We depict the graph of the total mass     as a function of $\alpha$  in Fig. 3 for
$\omega =0.15 $ (a) and  $\omega=0.5 $  (b) taking some different $k$ values.

Additionally, the gravitational surface redshift defined by $z := \frac{1}{f(r_b)} - 1$ is calculated as
 \begin{eqnarray}\label{z}
  z =\sqrt{\frac{3(\alpha-2)}{\alpha}} -1\;.
 \end{eqnarray}
We   see  that the gravitational redshift is independent of the parameters $\omega$ and $k$ then the limit  $\alpha \to \infty$ gives the upper bound for the redshift, $z < 0.732$. On the other hand, $\alpha=3$ gives $z=0$. Then $\alpha$ must be $\alpha>3 $ for the observational requirements. 
Variation of the surface redshift is shown in Fig. 4.

\begin{figure}[h]{}
	\centering
 { \includegraphics[width=0.5\textwidth]{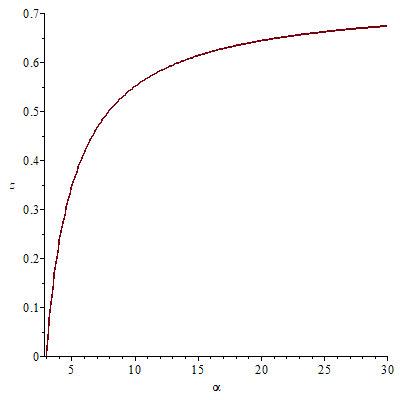}} 
	\\
	\parbox{6in}{\caption{{{\small{  The gravitational  surface redshift  versus the parameter $\alpha$. 
	}}}}}
\end{figure}

	In this model, we see that $\alpha$  can be fixed by the  gravitational surface redshift observations  due to (\ref{z}).
	However, the gravitational redshift  measurements do not have enough precise results \cite{Cottam2002,Lin2010}. Moreover, the observational value of the mass-radius ratio determines  one of the  two parameters  $\omega $ and $k$ from (\ref{totmass1}). If we can also predict  the total charge of star, we can fix $ \omega $ and $k$ separately from (\ref{Q2}).
 For example,   when we take  the gravitational redshift  of the neutron star EXO 0748-676 as $z=0.35$ with $M=2M_{\odot}$ and $R=13.1 km$ \cite{Lin2010}, we find  that  $\alpha\approx 5.08 $  from (\ref{z}) and there is one free parameter $k$ or $\omega$ that must be fixed in (\ref{totmass1}).  Then we can fix $k=1$  which leads to  $\omega \approx 0.01$.

\subsection{The Simple Model with $\alpha=4 $}
The model  simplifies   for $\alpha =4 $ as follows
\begin{eqnarray}\label{theory3}
L =    
\frac{1}{2\kappa} R*1
-    \epsilon_0\left[ 1 +     \frac{XR^2}{144 b\omega (k+1)}      \right]^{-\frac{ 9 k(\omega+1) }{X}} F\wedge *F  + 2A\wedge J + L_{mat}  + \lambda_a\wedge T^a\;. \ \ \ \ \ \ 
\end{eqnarray} 
where $X= 8k\omega +12k -10\omega -6$ for $\alpha=4$. 
Here we emphasize that the non-minimal function in this model can be expanded Maclaurin  series as
 \begin{eqnarray}
 Y(R) = 1- \frac{9k(\omega +1)}{144b \omega (k+1)} R^2 + \mathcal{O}(R^4)
 \end{eqnarray}
for $ |  \frac{XR^2}{144b\omega (k +1) } |   < 1$.
Then the model admits the interior metric 
 with the energy density, tangential pressure and electric field from  (\ref{rho1}), (\ref{pt1}), (\ref{E1}) and (\ref{metricin})
 
 \begin{eqnarray}\label{m2}
&& ds^2 _{in}= - a^2r^2 c^2 dt^2 + \frac{3}{1+br^4} dr^2 + r^2 d\Omega^2 \; \\
& &   \rho(r)=\frac{(k  + 1)[2- 2br^4   ]}{3\kappa c^2 r^2 (\omega +1)} \, , \label{rho2} \\
& &  p_t(r)=\frac{br^2 [2(k  +1) -X ]}{3\kappa  (\omega +1)}  + \frac{(k+1)(1- \omega)}{3\kappa (\omega+1) r^2}\, , \\
& &   E^2(r)= \frac{2\omega(k +1)}{3\kappa \epsilon_0  r^2 (\omega+1)} \left[1+\frac{X br^4}{4\omega(k  + 1)}\right]^{1+ \frac{9k (\omega+1)}{ X}} \,  .
\end{eqnarray}
 In this case, the model gives only one redshift which is $z= 0.225$ from (\ref{z}). 
Thus, if we observe the gravitational surface redshift for a compact star we can set the other parameters $k$ and $\omega$  for each observational mass value $M$ and boundary radius $r_b$  to describe the star.  

\subsection{The special case with  $k =1$}
Now we focus on the case with $k=1$ in which the equation of state must satisfy the special  constraint, $\rho c^2  = p_r+2p_t$ because of (\ref{trace}).
Now we compute the associated quantities by using the equations  (\ref{rho1}), (\ref{pt1}), (\ref{E1}) 
\begin{eqnarray}
& & \rho(r) = \frac{2[  2- br^\alpha (\alpha - 2) ]}{3\kappa c^2 r^2 (\omega +1 )} \, , \\
& & p_t(r) = \frac{(  1-\omega )[2- br^\alpha (\alpha - 2) ]}{3\kappa r^2 (\omega + 1)} \, , \label{pt} \\
& & E^2(r) = \frac{ 4\omega}{3\kappa \epsilon_0  r^2 (\omega + 1)} \left[ 1 + \frac{br^\alpha (3-\omega )(\alpha - 2)}{8\omega} \right]^{1-\frac{3(\omega + 1)(\alpha + 2)}{\alpha(\omega - 3)}} \, \label{Eofr}
\end{eqnarray}
with the interior metric (\ref{metricin}).
\begin{figure}[t]{}
	\centering
	\subfloat[ $\frac{\kappa Q^2}{16\pi^2\epsilon_0 r_b^2}  $ ]{ \includegraphics[width=0.5\textwidth]{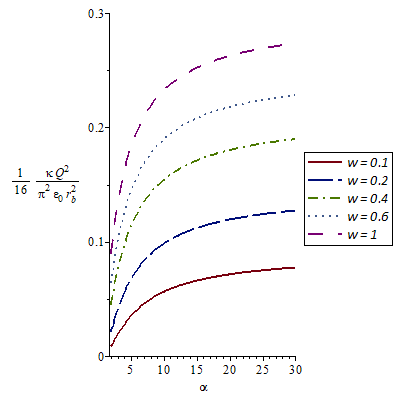} }
	\subfloat[ $\frac{GM}{c^2r_b}$  ]{ \includegraphics[width=0.5\textwidth]{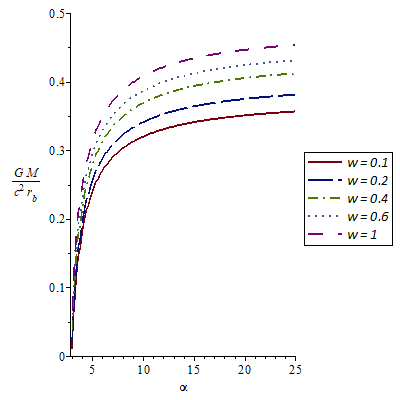}} 
	\\
	\parbox{6in}{\caption{{{\small{ The dimensionless quantities related with the total electric charge $Q$ (a)  and  the total mass $M$  (b)  as a function of  $\alpha$ for some different  $ \omega $  values and $k=1$.
	}}}}}
\end{figure}
The non-minimal coupling function (\ref{Y1}) becomes explicitly
\begin{eqnarray}
Y(R) =  \left[1 + \frac{b (3- \omega )(\alpha - 2)}{8\omega} \left(\frac{-R}{\alpha b + 2b}\right)^{\alpha/(\alpha - 2)}\right]^{\frac{3(\omega + 1)(\alpha + 2)}{\alpha(\omega - 3)}} \, .
\end{eqnarray}
 We also calculate the total charge  inside of the sphere with radius $r$ eliminating $b$ from (\ref{q})
\begin{eqnarray} \label{qin}
q^2(r) = \frac{64\pi^2\omega \epsilon_0 r^2}{3\kappa (\omega + 1)} \left[ 1 + \frac{br^\alpha (3- \omega )(\alpha - 2)}{8\omega}\right]^{\frac{4\omega \alpha + 6\omega + 6}{\alpha(\omega - 3)}} \, .
\end{eqnarray}
We check that the charge is regular at the origin for $\alpha  > 2$. 
 Then we can find the the total charge and mass  in terms of the boundary radius $r_b$, the parameters $\omega$ and $\alpha$ from  (\ref{Q2}) and (\ref{totmass1}).
\begin{eqnarray}
Q^2 = \frac{64\pi^2\epsilon_0\omega  r_b^2}{3\kappa (\omega + 1)}\left[1 + \frac{3-\omega }{4\omega} \right]^{\frac{4\omega \alpha + 6\omega + 6}{\alpha(\omega - 3)}},
\end{eqnarray}
\begin{eqnarray}
M = \frac{(\alpha - 3) }{(\alpha - 2)} \frac{c^2 r_b}{3G}+ \frac{2c^2\omega  r_b}{3G (\omega + 1)}\left[1 + \frac{3- \omega }{4\omega} \right]^{\frac{4\omega \alpha + 6\omega + 6}{\alpha(\omega - 3)}} \, . \label{totmass2}
\end{eqnarray}

 Variation of the total   mass and  electric charge as a function of the parameter $\alpha$ is shown in Fig. 5 for some different $\omega $ values. 
As we can see from the Figures that the increasing $\omega$ values increase the total mass and  electric charge. Then  we can find upper bound for the total mass and charge by  taking $ \omega =1$ and  $\alpha\rightarrow \infty$.
\begin{eqnarray}
\frac{GM}{c^2 r_b} = 0.48, \hskip 2 cm \frac{\kappa Q^2}{16\pi^2\epsilon_0 r_b^2} = 0.296
\end{eqnarray}

In this case, if we observe the gravitational surface redshift of a compact star, we can find $\alpha$ from (\ref{z}) and  $\omega$  from (\ref{totmass2}) for each observational mass  $M$ and boundary radius $r_b$ ratio $\frac{M}{r_b}$ to describe compact stars with the model. In future, the observations with  different redshifts in Table 1 lead to different $\omega$ and   $k$ values with the corresponding  $\alpha$.

\begin{table}[t]
	{\small 
		\centering
		\begin{tabular}{|c|c|c|c|c|c|}
			\hline  \ \
			Star   & $\frac{M}{r_b} \ (\frac{M_{\odot}}{km})$  &$\alpha$ &  $\frac{\kappa^2Q^2}{16\pi^2\epsilon_0 r_b^2}$ & 	$z$ (redshift) 	\\ \hline
			EXO 1745-248  & $\frac{1.4}{11} $ \cite{Ozel2009}
			&   3.940  & 0.054 & 0.215   \\
			\hline
			4U 1820-30  &    $\frac{1.58}{9.11}$ \cite{Guver2010} & 5.035 & 0.067 & 0.345  
			\\ \hline		
			4U1608-52		& $\frac{1.74}{9.3}$ 
			\cite{Guver20102}  
			& 5.611 &  0.073 & 0.390    \\
			\hline			
		
		\end{tabular}
		\caption{
			The dimensionless  parameter $\alpha $, the dimensionless  charge-radius ratio $\frac{\kappa^2Q^2}{16\pi^2 \epsilon_0 r_b^2} $ and   the surface redshift  $z$  obtained by using the observational mass $M$ and the radius $r_b$ for some neutron stars with $\omega =0.2 $ and $k=1 $ .	}
		\label{tab:template}
	}
\end{table} 
\section{Conclusion}

We have studied   spherically symmetric  anisotropic  solutions of  the non-minimally coupled $Y(R)F^2$ theory. We have established the non-minimal model which admits the regular interior metric solutions satisfying conformal symmetry  inside the star and Reissner-Nordstrom  solution at the exterior assuming the linear equation of state between the radial pressure and energy density as $p_r = \omega \rho c^2$. We found that the pressures and energy density decrease with the radial distance $r$ inside the star.

We matched the interior and exterior metric, and used the continuity conditions at the boundary of the star. Then we obtained such quantities as total mass, total charge and gravitational surface redshift in terms of the parameters of the model and the boundary radius of the star.
We see that the parameter $k$ can not  be  more than $1$ for non-negative tangential pressure and can not be less than $\frac{2}{\alpha} $ with  $ \alpha > 3  $  for the real valued   total mass and  electric charge.  The total mass and electric charge increases with  the increasing $\omega$ values, while the gravitational redshift does not change.  The total mass-boundary radius ratio has the upper bound $\frac{GM}{c^2r_b} = 0.48$ which is grater than the Buchdahl bound \cite{Buchdahl1959}   and the bounds given by \cite{Mak2001,Sert2017}.  The gravitational redshift at the surface only  depends on the parameter $\alpha$ and increases with increasing  $\alpha$ values up to the limit $z= 0.732$, which is the same result obtained from  the isotropic case with  $k=1$ \cite{Sert2017}.   We also investigated some sub cases such as $\alpha=4, k\neq 1$ and $\alpha>3, k=1$, which can be model of anisotropic stars.

We note that an interesting investigation of anisotropic compact stars  was recently  given by  Salako et al. \cite{Salako2018}     in the non-conservative theory of gravity. Then, the constants of the interior metric   were determined for some known   masses and radii. Then the physical parameters such as anisotropy,  gravitational redshift, matter density and pressures  were calculated and found that this model can describe even super-massive compact stars.
  On the other hand,  in our  conservative model (see  \cite{Sert2017} for energy-momentum conservation),   $\alpha$ can be determined by   the  gravitational surface redshift observations due to (\ref{z}).   Furthermore, if we can also determine  the mass-boundary radius ratio $\frac{M}{r_b}$ from observations,  we can fix one of the  two parameters  $\omega$ and $k$  via (\ref{totmass1}). Additionally, 
the observation of the total charge can fix both of the parameters  $\omega$ and $k$ from (\ref{Q2}). Thus we can construct a non-minimal $Y(R)F^2$ model for each charged compact star.
	But, the gravitational redshift  measurements  \cite{Cottam2002,Lin2010} and total charge predictions do not have enough precise results. However we can predict possible ranges of these quantities.

 \section*{Acknowledgement}

In this study the authors \"{O}.S. and F.\c{C}. were supported via the project number 2018FEBE001 and M.A. via the project number 2018HZDP036 by the Scientific Research Coordination Unit of Pamukkale University



\end{document}